\documentclass[%
 reprint,
 amsmath,amssymb,
 aps,
]{revtex4-1}

\usepackage{color}
\usepackage{graphicx}% Include figure files
\usepackage{dcolumn}% Align table columns on decimal point
\usepackage{bm}% bold math
\begin{document}

\preprint{APS/123-QED}

\title{The role of strangeness in hybrid stars and possible observables}

\author{V. Dexheimer}
\affiliation{Department of Physics, Kent State University, Kent OH 44242 USA}
\email{vdexheim@kent.edu}

\author{R. Negreiros}
\affiliation{Instituto de Fisica, Universidade Federal Fluminense, Niteroi, Brazil}

\author{S. Schramm}
\affiliation{FIAS, Johann Wolfgang Goethe University, Frankfurt, DE}

\date{\today}% It is always \today, today,

\begin{abstract}
We study the effects of strangeness on the quark sector of a hybrid star equation of state. Since the model we use to describe quarks is the same as the one we use to describe hadrons, we can also study the effects of strangeness on the chiral symmetry restoration and deconfinement phase transitions (first order or crossover). Finally, we analyze combined effects of hyperons and quarks on global properties of hybrid stars, like mass, radius and cooling profiles. It is found that a large amount of strangeness in the core is related to the generation of twin-star solutions, which can have the same mass as the lower or zero strangeness counterpart, but with smaller radii.
\end{abstract}

\pacs{Valid PACS appear here}% PACS, the Physics and Astronomy
\maketitle

%\tableofcontents

\section{\label{sec:level1}Introduction}

Recently, it has been understood that any realistic calculation for neutron star equations of state must take into account hyperons and/or quark degrees of freedom, since densities of several times nuclear saturation density can be reached in the center of the star. Moreover, many models that include hyperon and quark degrees of freedom are able to describe massive stars with masses around $2M_\odot$ (see \cite{Takatsuka:2004ch,Buballa:2014jta} and references therein). This means that the role played by strangeness at high densities has again become the focus of intense discussions. The interplay between the appearance of hyperons and quarks has also been discussed, as for example in Refs.~\cite{Zdunik:2012dj,Dexheimer:2012eu}.

In addition, recent studies of neutron star radii have indicated that these objects might have smaller radii than previously expected, around $10$ km or less \cite{Psaltis:2013fha,Guillot:2013wu}.  Although recent small radii measurements have been criticized (see for instance Refs.~\cite{Steiner:2010fz,Heinke:2014xaa}), the idea of small radii stars together with the constraint of stars with $2M_\odot$ \cite{Demorest:2010bx,Antoniadis:2013pzd} pushes toward equations of state very close to the causal limit, beyond which the speed of sound is larger than the speed of light. The so called ``hyperon puzzle", discussed in Ref~\cite{Zdunik:2012dj}, refers to the larger radii of massive stars containing hyperons. A possible solution, in this case, is the coexistence of strange hadronic and quark stars, which would separately fulfill mass and radius constraints. Such a possibility is achieved through another family of stars, referred to as ``twin stars" or tertiary stars \cite{Gerlach:1968zz,Kampfer:1981yr,Glendenning:1998ag,Schertler:2000xq,SchaffnerBielich:2002ki,Schramm:2013pma,Alvarez-Castillo:2013cxa,Blaschke:2013ana,Pagliara:2013gya,Pagliara:2014gja,Benic:2014jia}. These are stars containing quarks and with smaller radii than the respective stars containing hadrons. In this work, we investigate this possibility but in the context of stars with different strangeness content.

We make use of a self-consistent approach that includes hadrons and quarks in the same model to study the interplay between different degrees of freedom. We do that by changing the strength of the strange meson coupling to the quarks. This makes it straightforward to study the effect of the appearance of strangeness in neutron stars and allows, in addition, the study of the effect of different kinds of phase transitions in the system. Note that the effect of the strength of the strange meson couplings to baryons has recently been studied in detail, for example, in Refs.~\cite{Weissenborn:2011ut,Gusakov:2014ota,Lopes:2013cpa,2014arXiv1404.2428M}.
Within our framework, we study the possibility of smooth crossovers and first order phase transitions from hadronic dominant to quark dominant matter and from non-strange dominant to strange dominant matter. We then use our equation of state to obtain observables such as mass, radius and cooling profiles for neutron stars.

\section{\label{sec:level1}The Model}

Chiral sigma models are effective quantum relativistic models that describe hadrons interacting via meson exchange and, most importantly, are constructed from symmetry relations. They are constructed in a chirally invariant manner as the particle masses originate from interactions with the medium and, therefore, go to zero at high density and/or temperature. The nonlinear realization of the sigma model is an alternative approach to the widely used linear sigma model \cite{Papazoglou:1997uw,Lenaghan:2000ey,Nahrgang:2011mg} and it is in good agreement with nuclear physics results \cite{Papazoglou:1998vr,Bonanno:2008tt}.

The Lagrangian density of the SU(3) non-linear realization of the sigma model constrained further by astrophysics data can be found in Refs.~\cite{Dexheimer:2008ax,Schurhoff:2010ph}. A recent extension of this model which includes quarks as dynamical degrees of freedom \cite{Dexheimer:2009hi,2013PhRvC..88a4906H,Negreiros:2010hk} is described in the following. The Lagrangian density of the model in the mean field approximation reads
\begin{eqnarray}
L = L_{Kin}+L_{Int}+L_{Self}+L_{SB}\,,
\end{eqnarray}
where, besides the kinetic energy term for hadrons, quarks, and leptons, the terms:
\begin{eqnarray}
L_{Int}&=&-\sum_i \bar{\psi_i}[\gamma_0(g_{i\omega}\omega+g_{i\phi}\phi+g_{i\rho}\tau_3\rho)+M_i^*]\psi_i,\nonumber\\
L_{Self}&=&\frac{1}{2}(m_\omega^2\omega^2+m_\rho^2\rho^2+m_\phi^2\phi^2)\nonumber\\
&+&g_4\left(\omega^4+\rho^4+\alpha^2\frac{\phi^4}{2}+3\alpha(\omega^2+\rho^2)\phi^2\right)\nonumber\\&-&k_0(\sigma^2+\zeta^2+\delta^2)-k_1(\sigma^2+\zeta^2+\delta^2)^2\nonumber\\&-&k_2\left(\frac{\sigma^4}{2}+\frac{\delta^4}{2}
+3\sigma^2\delta^2+\zeta^4\right)
-k_3(\sigma^2-\delta^2)\zeta\nonumber\\&-&k_4\ \ \ln{\frac{(\sigma^2-\delta^2)\zeta}{\sigma_0^2\zeta_0}}\,,\nonumber\\
L_{SB}&=&-m_\pi^2 f_\pi\sigma-\left(\sqrt{2}m_k^ 2f_k-\frac{1}{\sqrt{2}}m_\pi^ 2 f_\pi\right)\zeta\,,
\end{eqnarray}
represent the interactions between baryons (and quarks) and vector and scalar mesons, the self interactions of scalar and vector mesons, and an explicit chiral symmetry breaking term (responsible for producing the masses of the pseudo-scalar mesons). The underlying flavor symmetry of the
model is SU(3) and the index $i$ denotes the baryon octet and the three light quarks. The mesons included are the vector-isoscalars $\omega$ and $\phi$ (strange quark-antiquark state), the vector-isovector $\rho$,
the scalar-isoscalars $\sigma$ and $\zeta$ (strange quark-antiquark state) and  the scalar-isovector $\delta$, with $\tau_3$ being twice the isospin projection of each particle. The isovector mesons affect isospin-asymmetric matter and are consequently important for neutron star physics. Note, that different self-interaction schemes for the vector mesons ($g_4$ term) can be included in the model, as long as they conform with chiral symmetry \cite{Dexheimer:2008ax}. The parameter $\alpha$ results from a renormalization of the vector fields in order to obtain their correct vacuum masses. Assuming equal contributions of the $\omega$ and $\rho$ meson, these constants can be absorbed in the coupling $g_4$ and only the $\phi$ field terms are affected (see Ref.~\cite{Papazoglou:1997uw} for a detailed discussion).

In the model presented in Ref.~\cite{Dexheimer:2009hi,2013PhRvC..88a4906H}, the degrees of freedom change due to the effective masses of the baryons and quarks. Here, we adopt a different formalism, explained in the following, and the effective masses for baryons and quarks are simply generated by the scalar mesons, except for a small explicit mass term $m_0$ 
\begin{eqnarray}
M_{B}^*&=&g_{B\sigma}\sigma+g_{B\delta}\tau_3\delta+g_{B\zeta}\zeta+m_{0_B}\,,\nonumber\\ 
M_{q}^*&=&g_{q\sigma}\sigma+g_{q\delta}\tau_3\delta+g_{q\zeta}\zeta+m_{0_q}\,.
\end{eqnarray}

The coupling constants of the model are shown in Table~\ref{tabela1}. They were fitted to reproduce the vacuum masses of the baryons and mesons, nuclear saturation properties (density $\rho_0=0.15$ fm$^{-3}$, binding energy per nucleon $B/A=-15.65$ MeV, nucleon effective mass $M^*_N=0.67$ $M_N$, compressibility $K=318.76$ MeV), asymmetry energy ($E_{sym}=32.43$ MeV with slope $L=102.77$ MeV), and reasonable values for the hyperon potentials ($U_\Lambda=-30.44$ MeV, $U_\Sigma=2.47$ MeV, $U_\Xi=-26.28$ MeV). The vacuum expectation values of the scalar mesons are constrained by reproducing the pion and kaon decay constants $f_\pi$ and $f_\kappa$.

The slope of the symmetry energy has become a very prominent constraint for the equation of state in the past years, as its measurements (through, for example, neutron skin experiments) have become more accurate. These results seem to indicate low values for this quantity ($L \sim 50-70$ MeV). See Refs.~\cite{Lattimer:2012xj,Tsang:2012se,Li:2013ola} or \cite{Horowitz:2014bja} for a recent review on the topic. Notice, however, that some works such as the one in Ref.~\cite{Cozma:2013sja} found much higher values for the slope of the symmetry energy ($L \sim 100$ MeV). A detailed work on the role of the symmetry energy in the Lagrangian density of the SU(3) non-linear realization of the sigma model is in progress and it will be available soon.

\begin{table}
\caption{\label{tabela1}Coupling constants for the model, using $\chi = 401.93$ MeV.}
\begin{ruledtabular}
\begin{tabular}{ccc}
$ g_{N\omega}=11.46 $&$ g_{N\rho}=3.83 $&$ g_{N\phi}=0 $ \\
$ g_{N\sigma}=-9.83 $&$ g_{N\delta}=0 $&$ g_{N\zeta}=1.22$ \\
$ g_{\Lambda\omega}=7.64 $&$ g_{\Lambda\rho}=0 $&$ g_{\Lambda\phi}=7.06 $ \\
$ g_{\Lambda\sigma}=-5.39 $&$ g_{\Lambda\delta}=0 $&$ g_{\Lambda\zeta}=-2.21$ \\
$ g_{\Sigma\omega}=7.64 $&$ g_{\Sigma\rho}=7.64 $&$ g_{\Sigma\phi}=7.06 $ \\
$ g_{\Sigma\sigma}=-3.88 $&$ g_{\Sigma\delta}=0 $&$ g_{\Sigma\zeta}=-4.36 $ \\
$ g_{\Xi\omega}=3.82 $&$ g_{\Xi\rho}=3.82 $&$ g_{\Xi\phi}=14.11 $ \\
$ g_{\Xi\sigma}=-1.54 $&$ g_{\Xi\delta}=0 $&$ g_{\Xi\zeta}=-7.66 $ \\
$ g_{u\omega}=g_{d\omega}=4.70 $& $ g_{u\rho}=g_{d\rho}=-2.00 $&$  g_{u\phi}=g_{d\phi}=0 $ \\
$ g_{u\sigma}=g_{d\sigma}=3.80 $&$ g_{u\delta}=g_{d\delta}=0 $&$ g_{u\zeta}=g_{d\zeta}=0 $ \\
$ g_{s\omega}=0 $& $ g_{s\rho}=0 $&$ g_{s\phi}=$variable \\
$ g_{s\sigma}=0 $&$ g_{s\delta}=0 $&$ g_{s\zeta}=-3.80 $ \\
$ g_4=38.5 $&$ k_0=1.19 \chi^2 $&$ k_1=-1.40 $ \\
$ k_2=5.55 $&$ k_3=2.66 \chi $&$ k_4=-0.07 \chi^4 $ \\
$ m_{0_u}=m_{0_d}=6$ MeV &$ m_{0_s}=72$ MeV &$ m_{0_N}=150$ MeV \\
$ m_{0_\Lambda}=376.58$ MeV &$ m_{0_\Sigma}=376.58$ MeV &$ m_{0_\Xi}=376.58$ MeV \\
\end{tabular}
\end{ruledtabular}
\end{table}

In order to suppress the hadrons at high density and/or temperature and allow the quarks to dominate, we introduce an excluded volume for the baryons. The use of such a technique was proposed long ago in 
Refs.~\cite{Baacke:1976jv,Hagedorn:1980kb,Gorenstein:1981fa,Kapusta:1982qd,Hagedorn:1982qh,Rischke:1991ke,Lattimer:1991nc,Cleymans:1992jz,Shen:1998gq,Bugaev:2000wz,Bugaev:2008zz,Satarov:2009zx,Hempel:2011kh} and recently used in 
Refs.~\cite{Steinheimer:2010ib,Steinheimer:2011ea,Dexheimer:2012eu}.  This is done by introducing the volume occupied by baryons and quarks as
\begin{eqnarray}
v_{B} = 0.64 {\rm \ fm}^{3} \ \ \ \ {\rm and} \ \ \ \  v_{q} = 0\,.
\end{eqnarray}
The $v_{B}$ value is chosen to represent the effect of the repulsive baryonic hard core with a corresponding radius of $r = 0.34$ fm. In this case, the chemical potential for baryons and quarks needs to be further modified in order to maintain thermodynamical consistency
\begin{eqnarray}
\hat{\mu_i}=\mu_i^*  - v_i P\,,
\end{eqnarray}
with
\begin{eqnarray}
\mu_i^*= Q_{B_i} \mu_B - Q_i \mu_Q - g_{i\omega} \omega - g_{\phi} \phi - g_{i\rho} \tau_3 \rho\,,
\end{eqnarray}
with $P$ being the total pressure of the system, $Q_{B_i}$ the baryon number of each particle, $\mu_B$ the baryon chemical potential, $Q_i$ the electric charge of each particle and $\mu_Q$ the charge chemical potential. In this way, the chemical potentials of the baryons are decreased by the appearance of quarks, but not vice versa. Furthermore, to be thermodynamically consistent, all particle densities, i.e., number density, energy density, entropy density, etc., have to be multiplied by a volume correction factor $f$, defined as the ratio of the unoccupied (excluded) volume $V'$ and the total volume $V$
\begin{eqnarray}
f = \frac{V'}{V} = \left(1 + \sum_i v_i Q_{B_i} \rho_i \right)^{-1}\,,
\end{eqnarray}
where $\rho_i$ is the density of each particle. In this way, the quarks effectively suppress the baryons by changing their chemical potential, while the quarks are only affected through the volume correction factor $f$.

\begin{figure}[t]
\vspace{3mm}
\includegraphics[width=8.7cm]{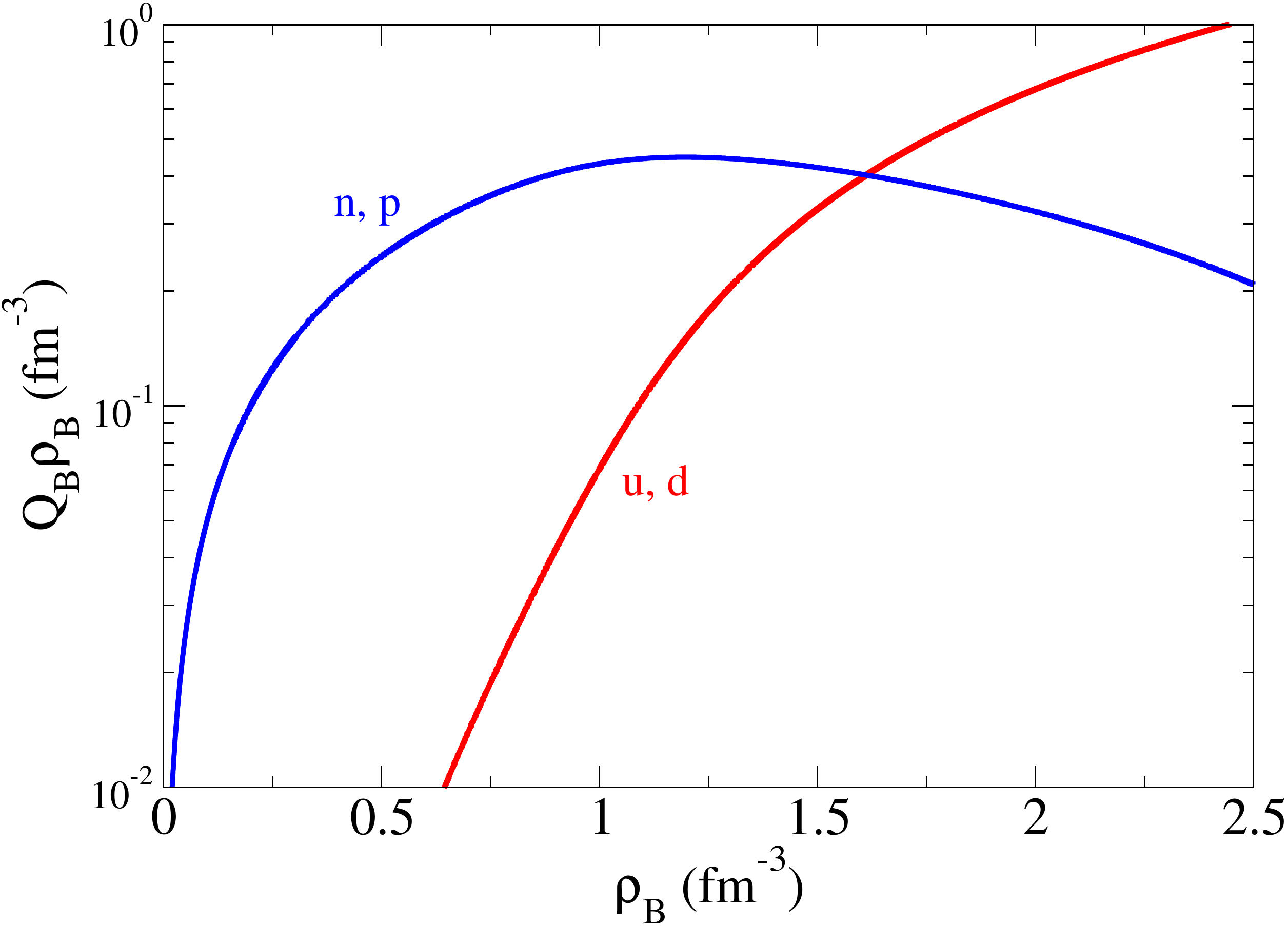}
\caption{(Color online) Population (particle density normalized by baryon number) as a function of baryon density for symmetric matter.\label{sum}}
\end{figure}

As a result, for symmetric matter, the model used in this work predicts that protons and neutrons are only suppressed at high densities, when a significant amount of quarks appear. This can be seen in Fig.~\ref{sum}. Due to the zero strangeness constraint, only the up and down quarks slowly appear at about $2.3$ times the saturation density. The strangeness constraint is necessary in order to compare theoretical results with nuclear and particle experimental results, which take place in a time interval much shorter than the weak equilibration time.

\section{\label{sec:level1}Stellar Matter and Structure}

\begin{figure}[t]
\vspace{3mm}
\includegraphics[width=8.2cm]{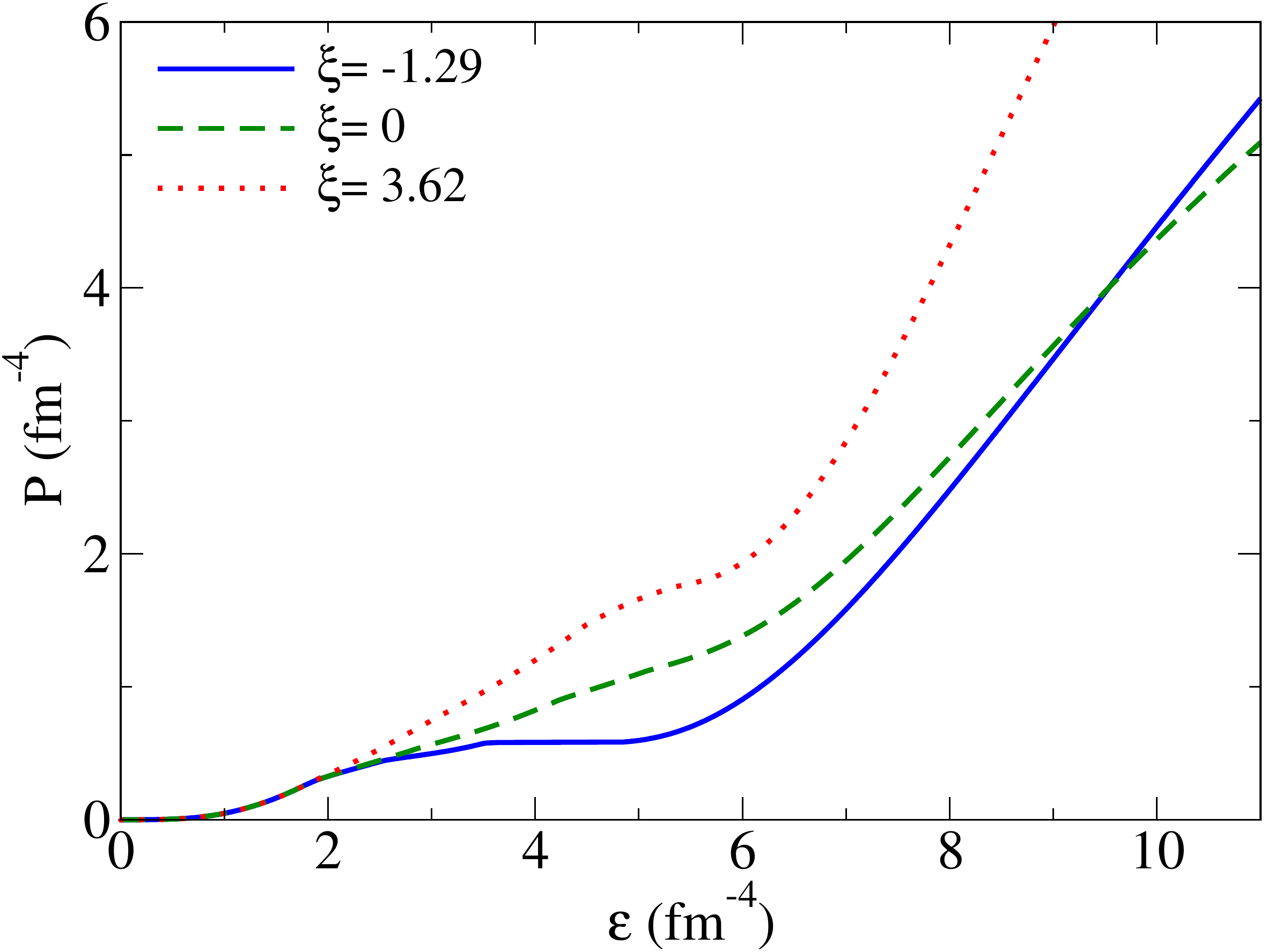}
\caption{(Color online) Equation of state for different strengths of the quark couplings.\label{eos}}
\end{figure}

In order to study neutron stars, we take into account charge neutrality and chemical equilibrium. Strangeness is not constrained since, for neutron stars, the time scale is large enough for strangeness not to be conserved. The equation of state for such a system is shown in Fig.~\ref{eos}. Depending on the strength of the quark couplings, the stiffness of the equations of state and the kind of phase transitions obtained are different. 

We show curves for different strengths of the quark coupling to the strange vector meson compared to the non-strange vector meson defined as $\xi=g_{q\phi}/g_{q\omega}$, keeping $g_{q\omega}$ constant.
Larger values of the parameter $\xi$ reproduce stiffer equations of state due to extra repulsion at intermediate and/or larger densities. This can be seen in Fig.~\ref{eos}. While most values of $\xi$ reproduce smooth equations of state, negative values below $\xi=-1.10$ reproduce first order phase transitions. So far, it is not understood what kind of deconfinement and chiral restoration phase transitions take place at small and zero temperatures. In Refs.~\cite{Baym:2008me,Lourenco:2012yv}, for example, it has been proposed that the phase transition in such a limit is a smooth crossover, as in the high temperature and low density regime.

\begin{figure}[t]
\vspace{3mm}
\includegraphics[width=9.5cm]{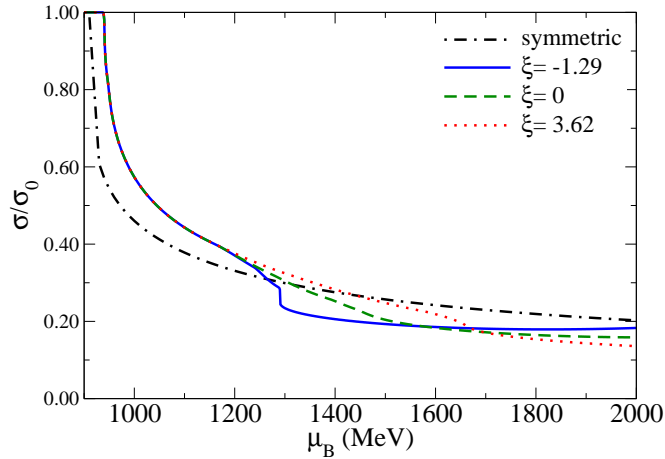}
\caption{(Color online) Scalar meson field (normalized by vacuum value) as a function of chemical potential for different strengths of the quark couplings.\label{sigma}}
\end{figure}

\begin{figure}[t]
\vspace{3mm}
\includegraphics[width=9.5cm]{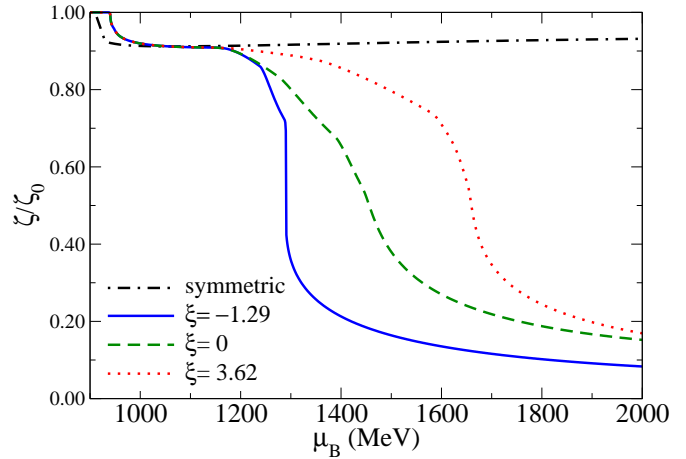}
\caption{(Color online) Strange scalar meson field (normalized by vacuum value) as a function of chemical potential for different strengths of the quark couplings.\label{zeta}}
\end{figure}

In order to understand better the results from Fig.~\ref{eos}, we show the variation of the scalar meson and the strange scalar meson fields for different strengths of the quark couplings in Figs.~\ref{sigma} and \ref{zeta}. The sigma field, usually referred to as the chiral condensate, is intrinsically related to the restoration of chiral symmetry and it is shown here also for the symmetric case. The zeta field is related to the quantity of strange particles in the system, meaning that a larger deviation from the vacuum value enhances the amount of hyperons (specially $\Xi$'s, which have larger strangeness) and strange quarks.

\begin{figure}[t]
\vspace{3mm}
\includegraphics[width=8.7cm]{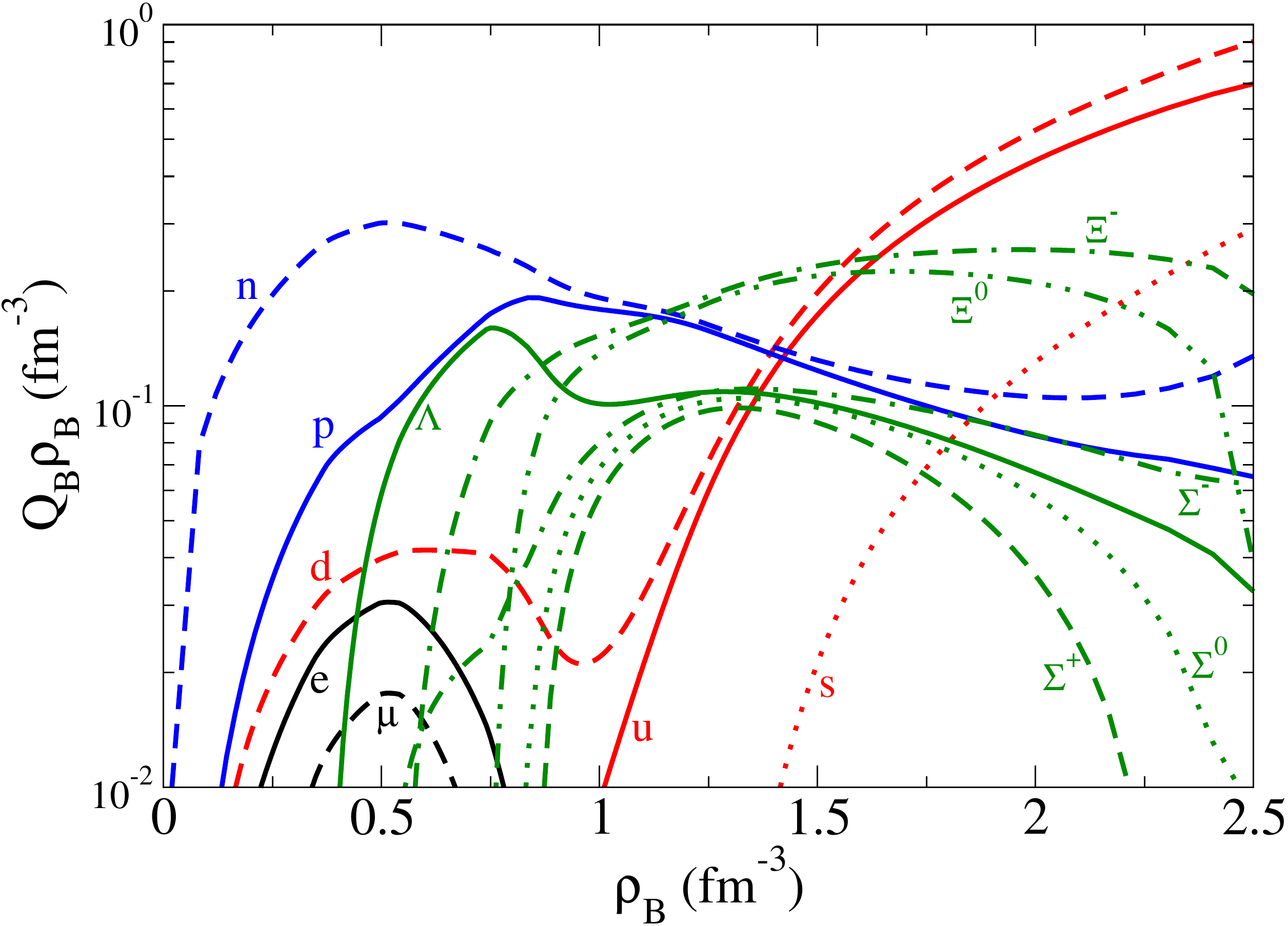}
\caption{(Color online) Population (particle density normalized by baryon number) as a function of baryon density for $\xi=3.62$.\label{pop1}}
\end{figure}

Now, we discuss the particle population present in the stars reproduced by the parametrizations discussed above. For $\xi=3.62$, Fig.~\ref{pop1} shows that the down quark appears very early in the system, followed by the hyperons $\Lambda$, $\Sigma^-$, $\Xi^-$, $\Xi^0$, $\Sigma^0$ and $\Sigma^+$, followed by the up and strange quarks.  As the density increases, the down quark is temporarily suppressed by the appearance of hyperons, but it becomes dominant at high densities. The strange quark only appears at very high densities, as it not only has a larger bare mass, but is also suppressed by the strong positive coupling to the strange vector meson.

Note that a sequential occurrence of deconfined quarks with different flavors at high densities has already been discussed in Ref.~\cite{Blaschke:2008gd}. In that work, the chiral symmetry restoration (associated with the quark deconfinement) is calculated in the NJL model by solving the gap equations and taking into account charge neutrality in such a way that, similar to our case, it depends on the quark chemical potentials. These chemical potentials are in turn dependent on the electric charge and bare mass of particles. In this scenario, the down quark drips out of the baryons first, followed by the up and strange quarks. At high temperature and zero chemical potential, sequential quark deconfinement has also been seen in lattice QCD calculations \cite{Bellwied:2013cta}.

\begin{figure}[t]
\vspace{3mm}
\includegraphics[width=8.7cm]{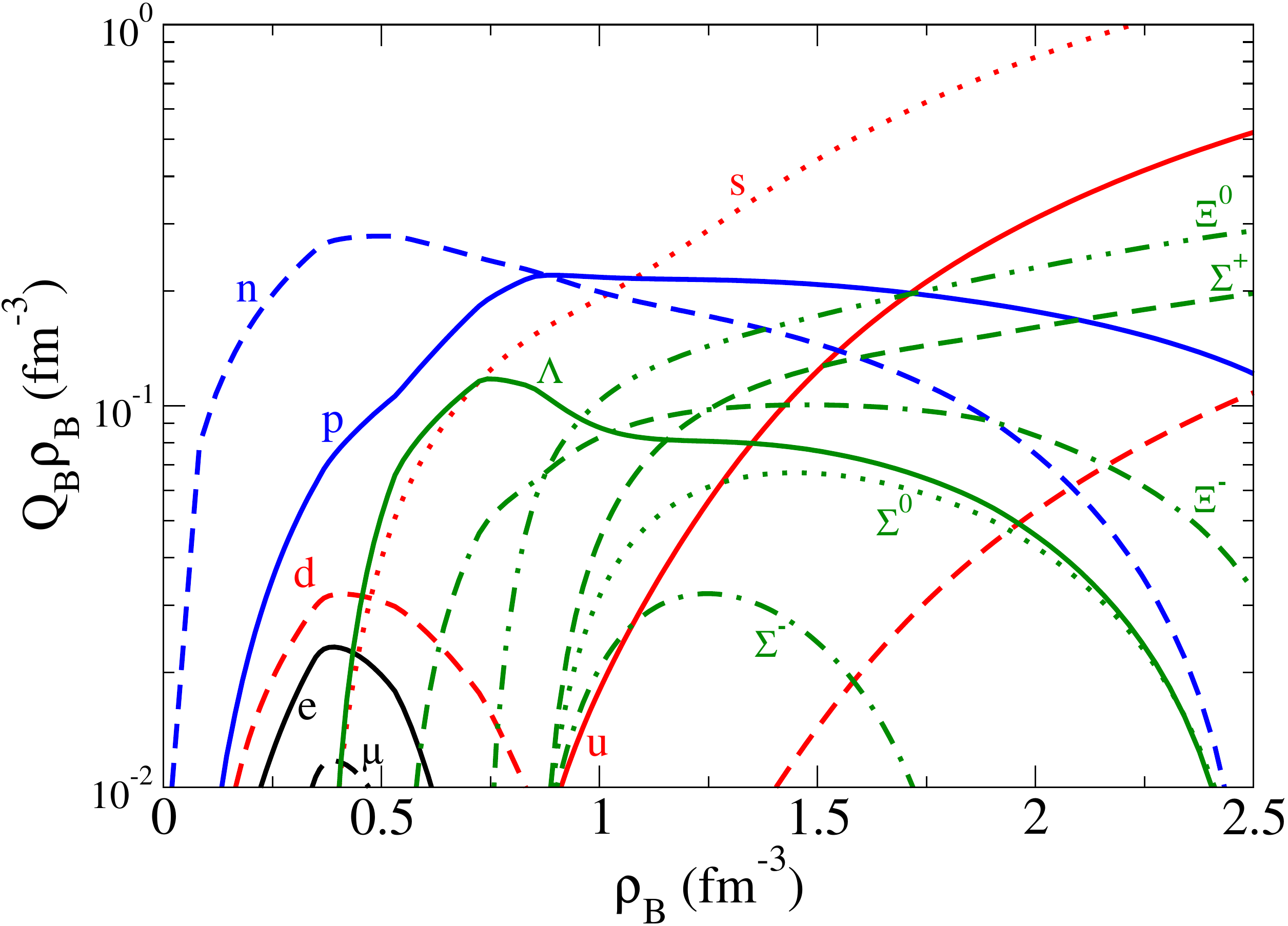}
\caption{(Color online) Population (particle density normalized by baryon number) as a function of baryon density for $\xi=0$.\label{pop2}}
\end{figure}

Fig.~\ref{pop2} shows that the down quark appears once more very early in the system for $\xi=0$, followed by the hyperon $\Lambda$ together with the strange quark, and then the hyperons $\Xi^-$, $\Xi^0$, $\Sigma^+$, $\Sigma^0$ and $\Sigma^-$, followed by the up quarks. Note that the early presence of the strange quarks in this case (without the influence of the $\phi$ meson) suppresses the negative charged hyperons, which would normally appear before the others to fulfill charge neutrality (now taken care of by the strange quark).

Note also, that we could have used any positive value for the parameter $\xi$. The choices of zero and $\xi=3.62$ were only made to show the largest  possible span of effects. After $\xi=3.62$, the effect of different $\phi$ coupling with the strange quark saturates, not showing further changes. This is, because at $\xi=3.62$, the strange quarks appear after all of the other possible particles, and in a small amount.

For negative values of the $\xi$ parameter, there is no change in the order of the appearance of the quarks and hyperons with respect to density, since the strange quark appears basically at the same low density as in the  $\xi=0$ case. Nevertheless, for negative enough values of the $\xi$ parameter ($\xi<-1.10$), the transition from hadronic-dominant to quark-dominant matter, and the transition from non-strange-dominant to strange-dominant matter (which so far have been smooth crossovers) become a first order phase transition.  

\begin{figure}[t]
\vspace{3mm}
\includegraphics[width=8.7cm]{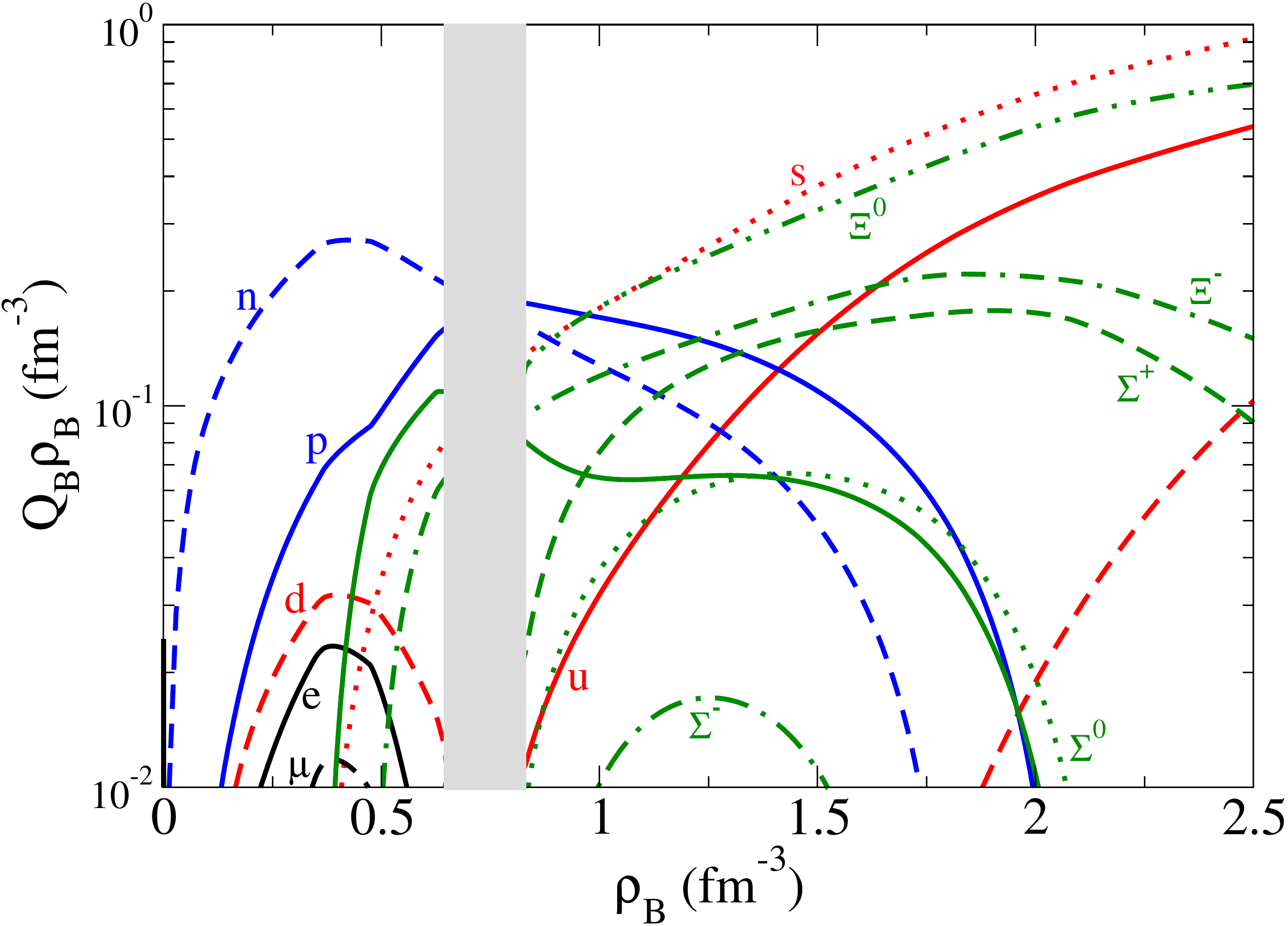}
\caption{(Color online) Population (particle density normalized by baryon number) as a function of baryon density for $\xi=-1.29$.\label{pop3}}
\end{figure}

This had already been shown in Fig.~\ref{eos} but can be better understood in Fig.~\ref{pop3} plotted for $\xi=-1.29$, where a jump in the baryon density (which is a first derivative of the baryon chemical potential) can be observed. The grey shaded area in the figure represents an unphysical region, which would collapse to a point in the presence of gravity, as it corresponds to a single value of pressure. 

\begin{figure}[t]
\vspace{3mm}
\includegraphics[width=9.6cm,clip,trim= 0 0 0 75]{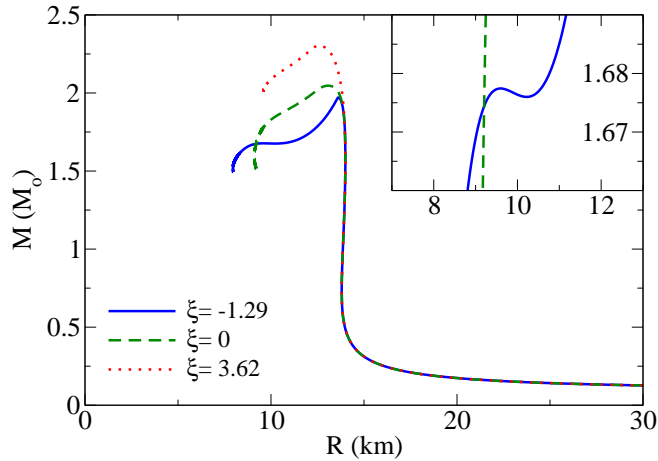}
\caption{(Color online) Mass-radius diagram considering different strengths of the quark couplings. The inset highlights the twin star configurations.}\label{tov}
\end{figure}

The equations of state for the three cases already discussed are analyzed under the influence of gravity in Fig.~\ref{tov}, from the solutions of the Tolman-Oppenheimer-Volkoff equations \cite{Tolman:1939jz,Oppenheimer:1939ne}. When  $\xi=3.62$, a maximum mass of $2.31 M_\odot$ is obtained with a corresponding radius of  $12.58$ km. When  $\xi=0$, a maximum mass of $2.05 M_\odot$ is obtained with a corresponding radius of  $13.07$ km. Finally, when $\xi=-1.29$, a maximum mass of $1.97 M_\odot$ is obtained with a corresponding radius of  $13.61$ km. 

An interesting feature of negative values of the $\xi$ parameter is that a secondary family of stars appears (at $\xi=-1.26$) with significantly smaller radii. This feature is related to the baryon density discontinuity at the phase transition and it is often called ``twin stars solution". These stars have equivalent masses (to the normal branch), but different radii. 

As $\xi$ becomes more negative, the amount of stars in the twin family increases. When $\xi=-1.29$, the first order phase transition to dominant quark strange matter reproduces twin stars with radii spanning  from $9.60$ to $10.22$ km and a maximum mass  of $1.68 M_\odot$. The discontinuity, across the phase transition that allows for such a configuration is quite small, $\Delta\rho_B=0.18$ fm$^{-3}$ (equivalent to an energy density discontinuity of  $\Delta\epsilon=238.40$ MeV/fm$^3$). 

Note that the star branch with smaller radii in Fig.~\ref{tov}, with hyperon and quark contributions, has significantly more strangeness than the normal branch. Observation of such configurations would point to the confirmation of first order phase transitions in stars, as already pointed out by Ref.~\cite{Glendenning:1998ag,Schertler:2000xq,Alvarez-Castillo:2013cxa,Benic:2014jia}. In the context of hadronic matter, the possibility of smooth and strong phase transitions to strange matter has been explored in Ref.~\cite{Gulminelli:2013qma}. For $\xi=-1.29$, for example, a star mass of $1.68 M_\odot$, corresponding to radii 
of  $14.00$ km and $9.60$ km in different branches, contains strangeness of $f_s = 0.01$ and $f_s = 1.68$, respectively, at the center. The total strangeness is defined as $f_s = \sum_i \rho(i) Q_{S_i} / \rho_B$, where $Q_{S_i}$ is the strangeness of each particle. For values of $\xi$ more negative than $-1.29$, the twin star solutions exist, but their equations of state becomes supraluminal for high enough densities. Therefore, we will not discuss them in this work. For a complete review on the topic of the behavior of models containing the excluded volume technique at high densities, see Ref.~\cite{Satarov:2014voa}. For a review on the topic of the sound speed in neutron stars see Ref.~\cite{Bedaque:2014sqa}.

\begin{figure}[t]
\vspace{3mm}
\includegraphics[width=9.7cm,clip,trim=100 70 0 57]{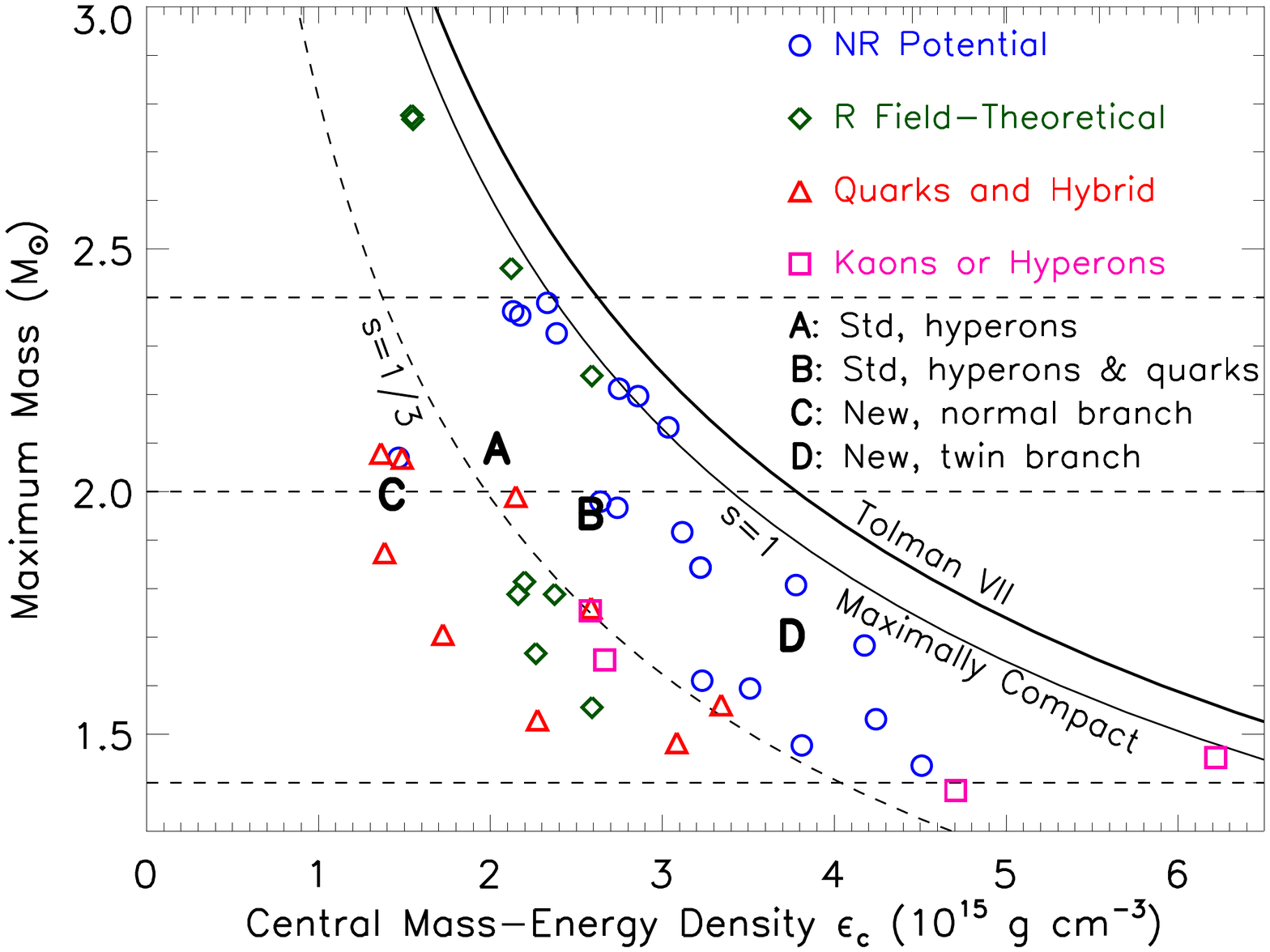}
\caption{(Color online) Diagram showing different models for neutron stars placed with respect to their compactness. Figure adapted from Refs.~\cite{Lattimer:2004sa,Lattimer:2012nd} by Lattimer and Prakash.\label{lattimer}}
\end{figure}

Figure~\ref{lattimer} (modified from Refs.~\cite{Lattimer:2004sa,Lattimer:2012nd}) shows the relation between the maximum mass star and the corresponding central density predicted by several classes of models from the literature, such as nonrelativistic (NR) potential or Skyrme-like, relativistic (R) field theoretical, EOS's containing significant amounts of quarks, and EOS's with significant contributions from hyperons or meson condensates. Note that the EOS's from Refs.~\cite{Dexheimer:2008ax,Dexheimer:2009hi}, which derive from another version of the model discussed in this work, are also shown marked by the letters ``A" and ``B". Finally, we analyze the compactness of the equation of state generated from the parametrization $\xi=-1.29$ discussed in this work (marked by ``C" and ``D" in the figure). The letter ``C" denotes the maximum mass star of the normal branch, and the letter ``D" the one of the twin branch. Note that ``D" is beyond the limiting curve ``s=1/3", which delimits the supraluminal behavior for quark stars when modeled by the simple MIT bag model. It is interesting to note that this line is also a limit to many models which reproduce a substantial amount of quarks in stars. The distance between ``C" and ``D" points to the difference between the compactness of both star families reproduced (normal and twin).

In addition, we would like to compare our results with Ref.~\cite{Alford:2013aca} by Alford and Han, which discusses different categories of twin stars assuming a Maxwell construction between different phases of matter and a constant speed of sound. Although our case with $\xi=-1.29$ reproduces twin star solutions, it shows no connected branch (stars with density beyond the phase transition jump but still in the stable part of the main star branch in the mass-radius diagram) within our numerical accuracy. We believe that we might reproduce a connected branch in agreement with Ref.~\cite{Alford:2013aca}, however, in practice, it may be too small to be seen. We reproduce a ratio of $\Delta\epsilon$/$\epsilon_{\rm trans}=0.33$, where $\Delta\epsilon$ is the jump in energy density across the phase transition and $\epsilon_{\rm trans}$ the energy density at which the transition takes place, and we reproduce a ratio $P_{\rm trans}$/$\epsilon_{\rm trans}=0.16$, where $P_{\rm trans}$ is the pressure at which the transition takes place. These values are important as they are connected to the ability of the low density phase to be supported by a large enough high density phase in the core. An intermediate size for the high density phase in the core necessarily turns the star unstable, originating disconnected solutions for stars. In other words, the existence of the twin branch depends on the size of the discontinuity across the phase transition and the stiffness of the equations of state of both phases.

\section{\label{sec:level1}Thermal Evolution}

We now turn our attention to the thermal evolution of the objects whose structures and compositions were discussed above.
The thermal behavior of a compact star strongly depends on its microscopic and macroscopic properties, thus, when combined with observational data, it is a formidable way of probing the characteristics of these objects. 
The thermal evolution of a compact star is governed by the general relativistic thermal balance and transport equations, given by ($G = c = 1$) \cite{Weber}
\begin{eqnarray}
  \frac{ \partial (l e^{2\phi})}{\partial m}& = 
  &-\frac{1}{\rho \sqrt{1 - 2m/r}} \left( \epsilon_\nu 
    e^{2\phi} + c_v \frac{\partial (T e^\phi) }{\partial t} \right) \, , 
  \label{coeq1}  \\
  \frac{\partial (T e^\phi)}{\partial m} &=& - 
  \frac{(l e^{\phi})}{16 \pi^2 r^4 \kappa \rho \sqrt{1 - 2m/r}} 
  \label{coeq2} 
  \, ,
\end{eqnarray}
where the macroscopic dependence is given by the variables $r$, $\rho(r)$ and $m(r)$, that represent the radial distance, the
energy density, and the stellar mass, respectively. The thermal properties are represented by the temperature $T(r,t)$, luminosity $l(r,t)$, neutrino emissivity $\epsilon_\nu(r,T)$, thermal conductivity
$\kappa(r,T)$ and specific heat $c_v(r,T)$. Furthermore, the boundary conditions of Eqs.~(\ref{coeq1}) and (\ref{coeq2}) are provided by the vanishing heat flux at the center of the star and the luminosity at its surface, defined by the relationship between the mantle and photosphere temperature \cite{Gudmundsson1982,Gudmundsson1983,Page2006}. 

\begin{figure}[t]
\vspace{3mm}
\includegraphics[width=9.cm,clip,trim=12 0 0 0]{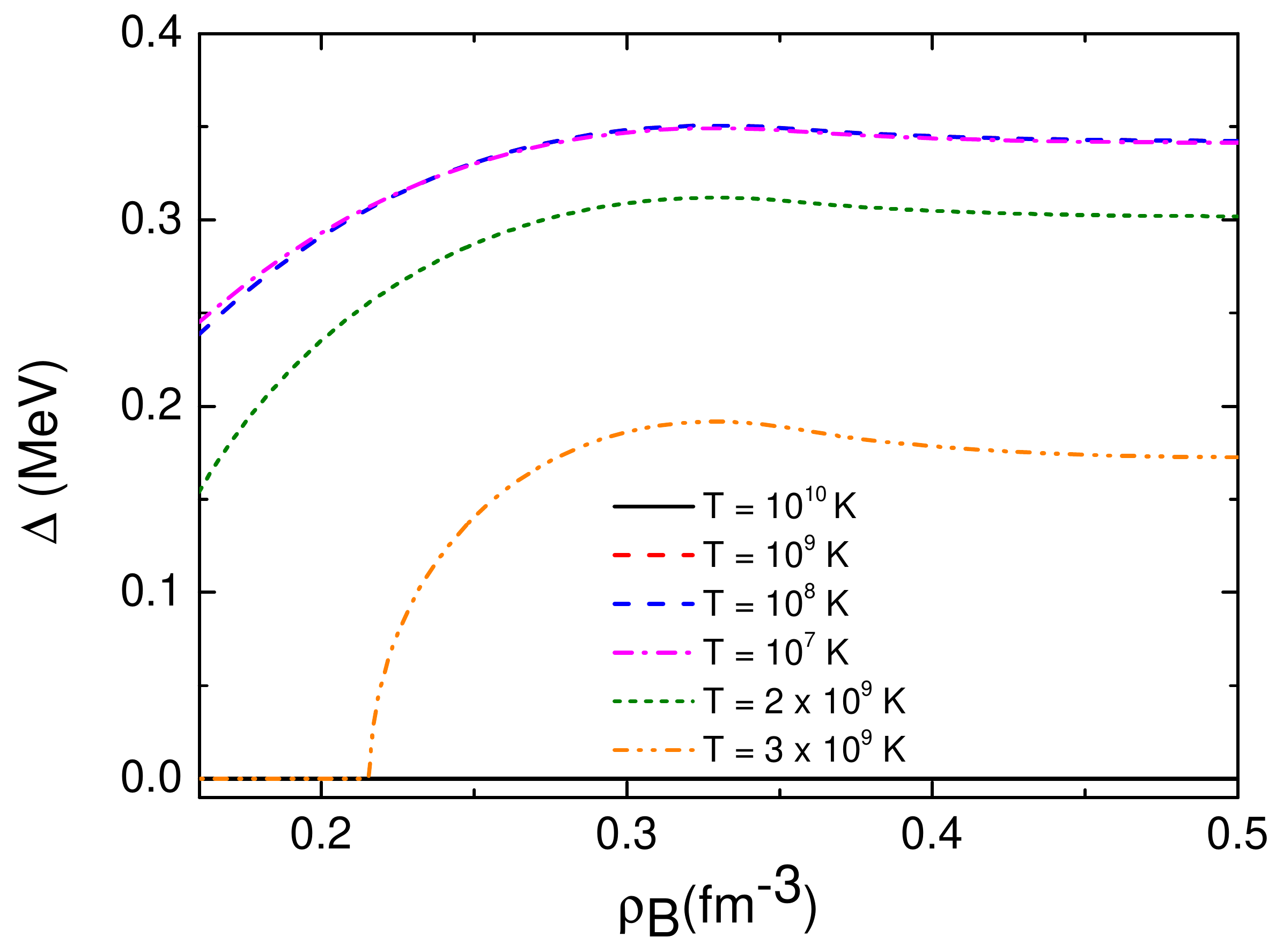}
\caption{(Color online) Gap energy for the neutron triplet pairing shown for different temperatures. \label{Delta}}
\end{figure}

In our study, we consider all state of the art neutrino emission processes for the thermal evolution. In the hadronic phase,
we take into account the direct Urca, modified Urca and bremsstrahlung processes, as well as the Pair-Breaking-Formation process (PBF) that accompanies pairing (described below) \cite{Yakovlev2001a}.  For the quarks we consider the quark direct Urca, quark modified Urca, and quark bremsstrahlung processes.
Furthermore, we consider the possibility of hadronic as well as quark pairing. For the hadronic model, we consider proton singlet pairing ($^1S_0$), neutron triplet pairing ($^3P_2$) for the core region and neutron singlet ($^1S_0$) pairing for the crustal region. To illustrate gap values, we show in Fig.~\ref{Delta} the neutron triplet gap energy as a function of density, for several values of temperature. Note that the proton pairing has a similar behavior, except for much stronger pairing, which is necessary to explain the behavior of Cas A \cite{Page2011a,Yakovlev2011}. For the quarks, we consider pairing with a gap $\Delta = 10$ MeV.

\begin{figure}[t]
\vspace{3mm}
\includegraphics[width=10.2cm,clip,trim=45 0 0 48]{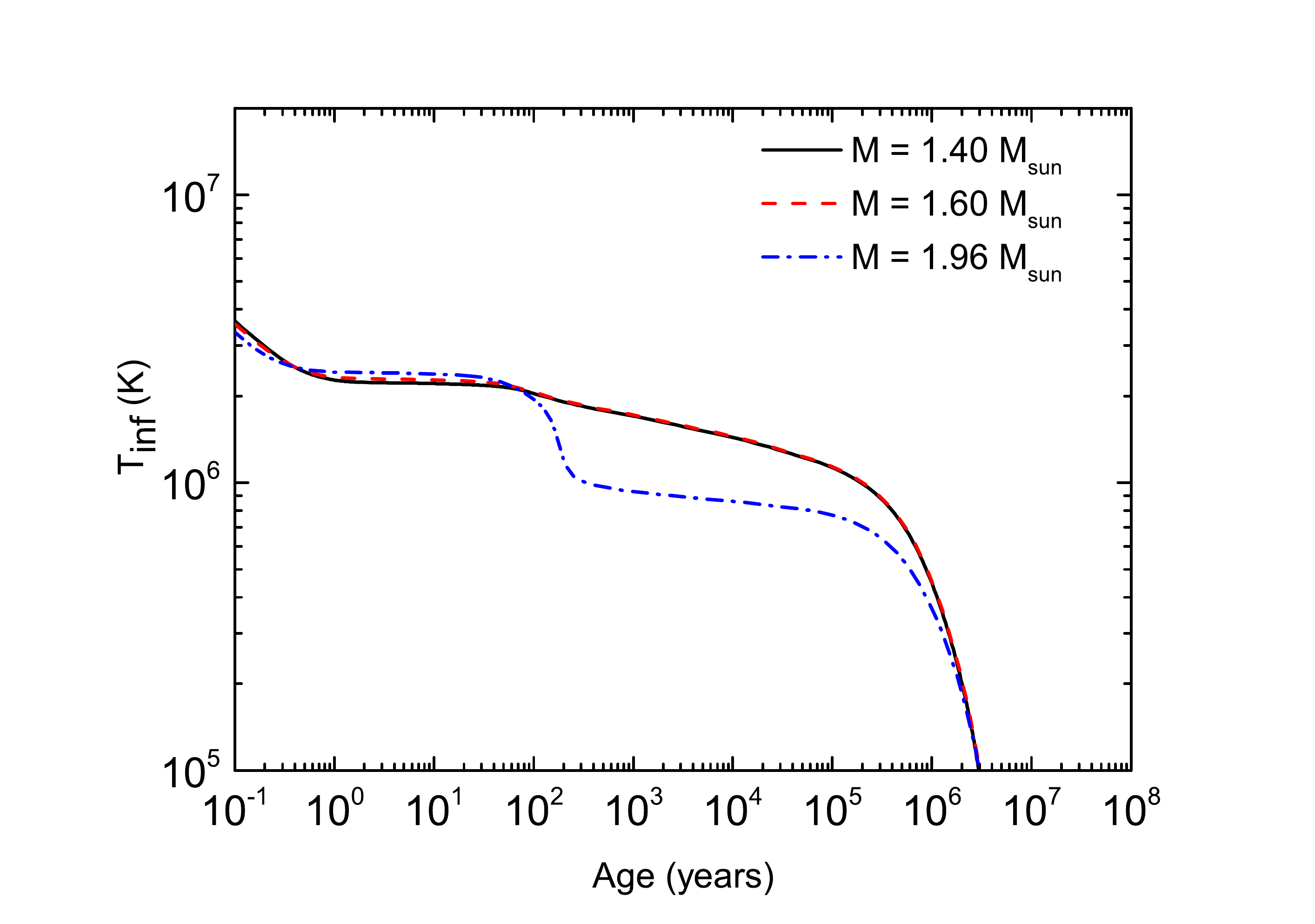}
\caption{(Color online) Thermal evolution (red-shifted surface temperature as a function of time) of stars for $\xi=-1.29$.\label{cool1}}
\end{figure}

\begin{figure}[t]
\vspace{3mm}
\includegraphics[width=10.2cm,clip,trim=45 0 0 0]{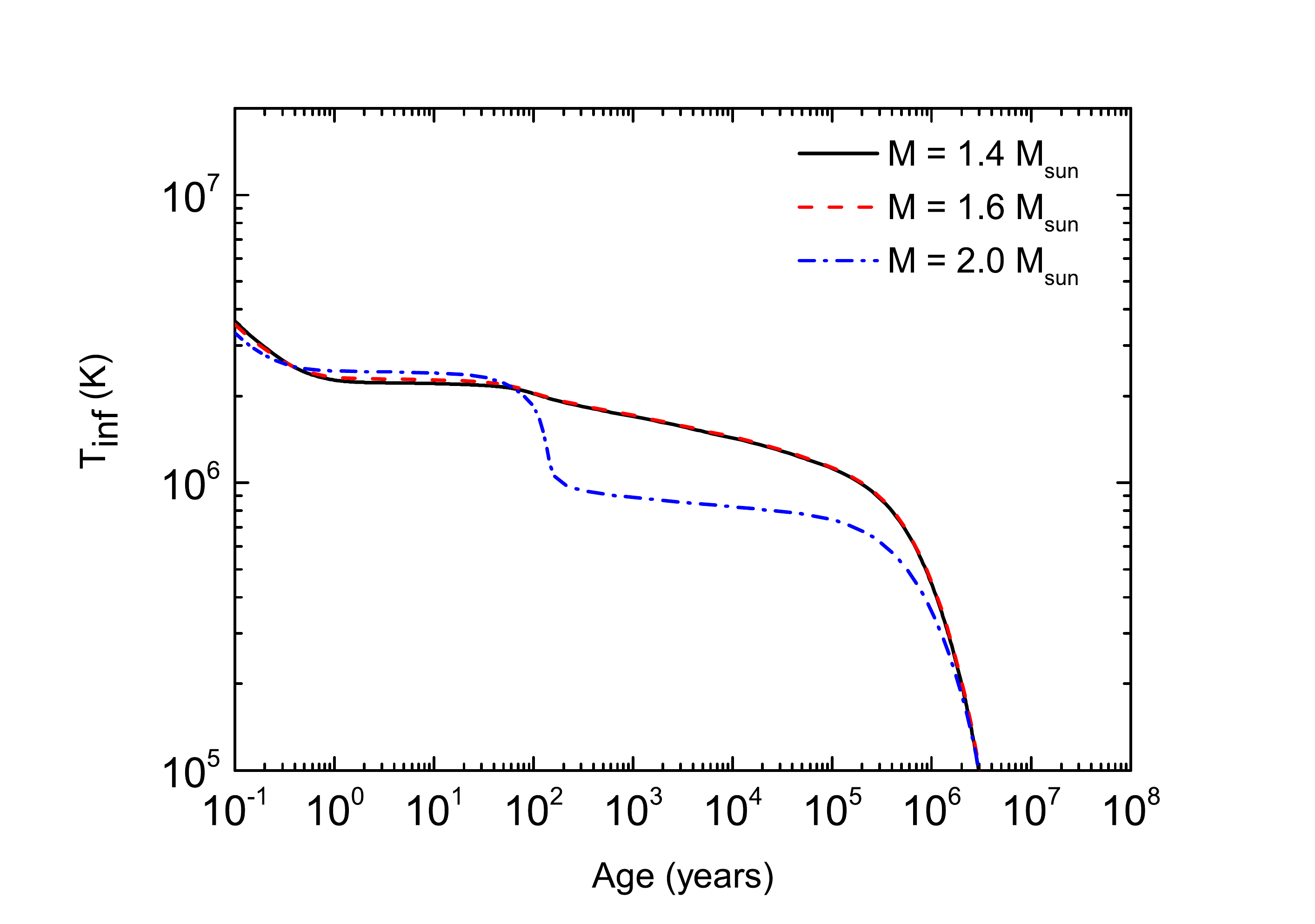}
\caption{(Color online) Thermal evolution (red-shifted surface temperature as a function of time) of stars for $\xi=0$.\label{cool2}}
\end{figure}

\begin{figure}[t]
\vspace{3mm}
\includegraphics[width=10.2cm,clip,trim=45 0 0 0]{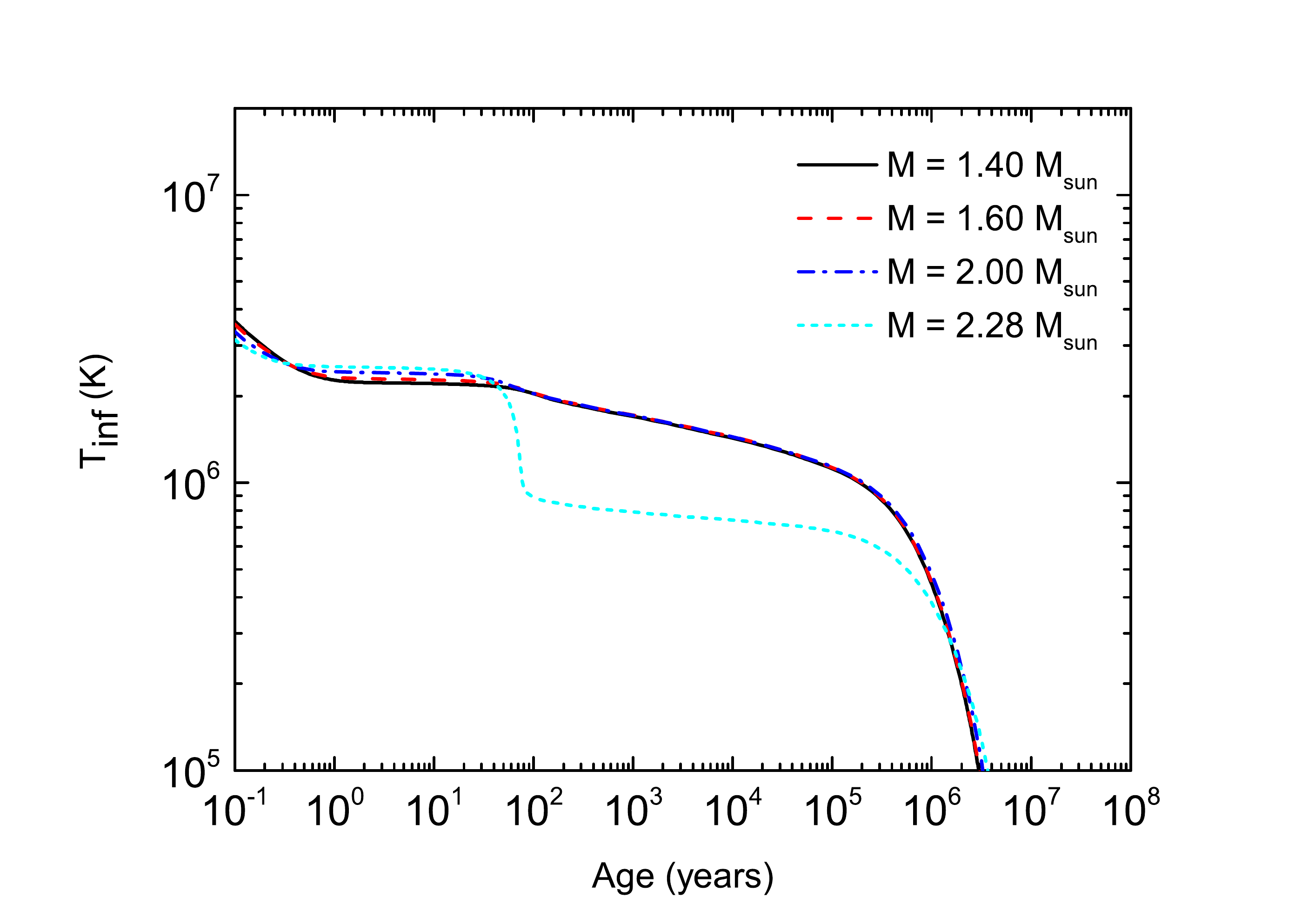}
\caption{(Color online) Thermal evolution (red-shifted surface temperature as a function of time) of stars for $\xi=3.62$.\label{cool3}}
\end{figure}

We start by showing the results for the thermal evolution of stars generated using the parameter $\xi=-1.29$ for the strength of the quark coupling, which we display in Fig.~\ref{cool1}. In the case of neutron stars of relatively low masses (1.4 and 1.6 $M_\odot$) their composition is similar enough that their thermal evolution is almost indistinguishable. As the star masses increase, however, the thermal evolution starts to exhibit a faster behavior, as expected in this case. 
Such a qualitative behavior is also exhibited in the thermal evolution of stars generated using the parameters $\xi=0$ and $\xi=3.62$, as shown in Figs.~\ref{cool2} and \ref{cool3}.

\begin{figure}[t]
\vspace{3mm}
\includegraphics[width=10.2cm,clip,trim=45 0 0 0]{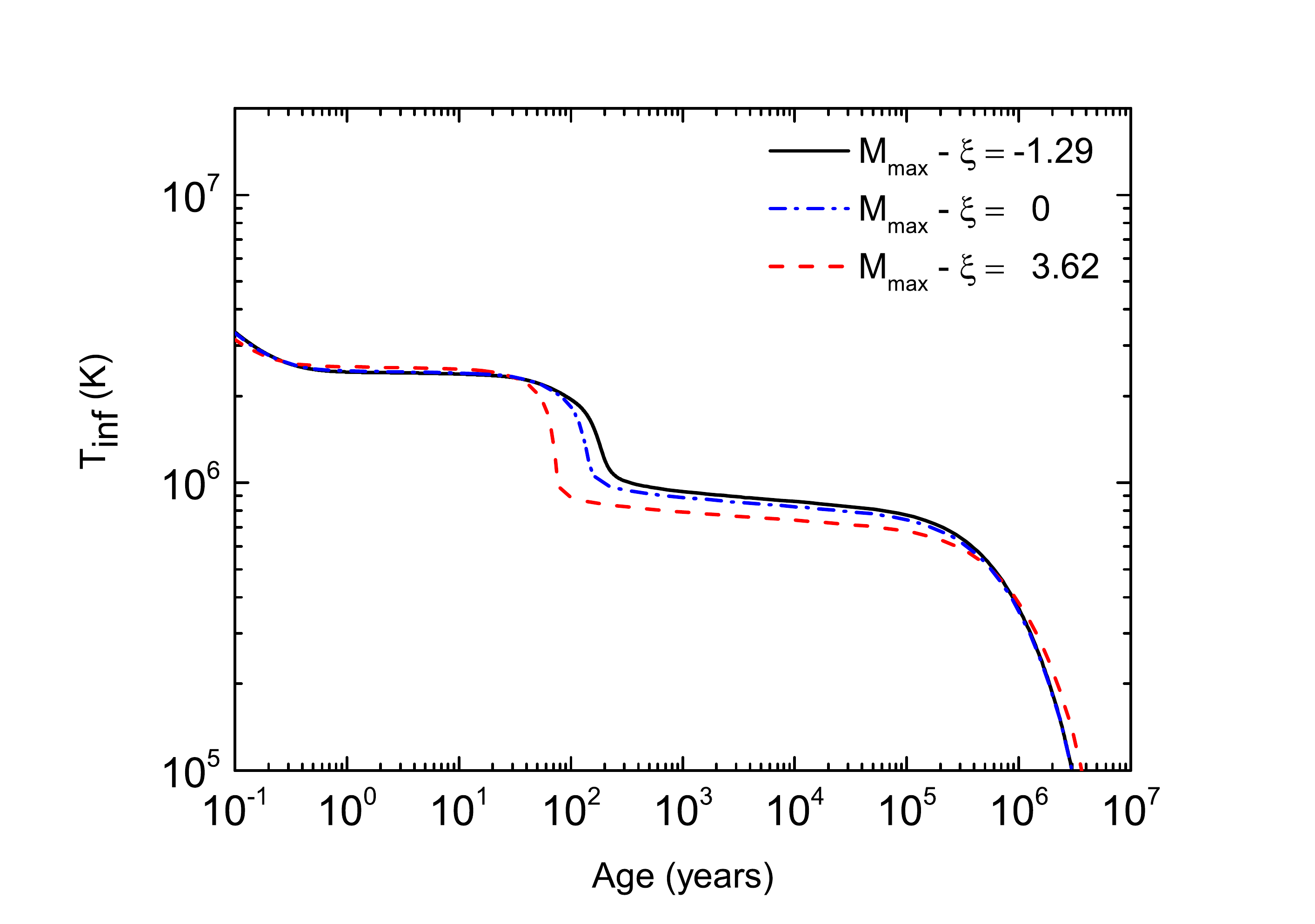}
\caption{(Color online) Thermal evolution of the maximum mass star from Figures 11, 12 and 13.\label{cool4}}
\end{figure}

The similarity among the thermal evolution of different model parametrizations of the quark coupling to the strange vector meson compared to the non-strange vector meson $\xi$ is not surprising, in particular for the low mass stars, where quark matter is not strongly present. A possible way of differentiating between the parametrizations studied is to investigate the cooling of the maximum mass star for each of them. These objects have the highest possible quark matter content (in the normal branch) allowed by each parametrization and, thus, should exhibit differences in their cooling. This is shown in Fig.~\ref{cool4}. Note that, in this way, we can also obtain information about the strangeness (through the mass of the star) in the cooling curves. Although we do not have observational information regarding the mass of isolated stars, we could at the very least infer some constraints by using the fact that 
 we know that larger strangeness implies slower star cooling, among other things, due to the smaller underlying stellar mass.

As shown in Fig.~\ref{cool4}, although all parametrizations exhibit the same qualitative behavior, the parametrization $\xi=-1.29$ seems to be slightly better if one is to interpret recent thermal observations such as in Cas A. The core-crust thermal coupling in this case (signaled by the sudden drop in surface temperature) takes place a later ages (when compared to the other parametrizations). This might facilitate the interpretation of Cas A data as this object exhibits a surface temperature drop at around $\sim 300$ years. In practice, there are other factors in play, in particular the pattern of the superconducting/superfluidity phases. A thorough investigation of this subject will be performed in a future investigation.

\section{\label{sec:level1}Conclusions}

We studied the dependance of neutron star properties on the strength of the quark couplings (strange vector to vector quark coupling ratio). The choice of values for this quantity affects the stiffness of the equation of state and the strength of the phase transition to (dominantly) deconfined strange matter, ranging from crossovers to first order phase transitions. We performed this task in a controlled way by making use of a self-consistent model that includes hadrons and quarks degrees of freedom. In this model, the interactions determine the density at which deconfinement and chiral symmetry restoration take place and the inclusion of an excluded volume for the hadrons ensures that they are not present at high densities.

The effect of strangeness in neutron stars emerges in the mass-radius relation, where a large amount of strangeness is related to the generation of twin-stars, which can have the same mass as the lower or zero strangeness counterpart, but with smaller radii.  The measurement of such stars would be a clear indication of a first order phase transition taking place at high densities and low temperatures, at least for charge neutral and beta-equilibrated matter. Nevertheless, this would provide us a priceless insight in the understanding of the QCD phase diagram.

With respect to the thermal evolution of stars, we have shown that the three quark coupling parametrizations studied present distinct behavior for massive stars, where the quark phase manifests itself differently for the three cases investigated. All three parametrizations, however, seem to be in agreement with the current cooling picture, in which high mass objects present a relatively faster thermal evolution than those objects with less mass and pairing in the hadronic and quark phases is necessary if one is to agree with recent data such as that obtained for Cassiopeia A. Among the three parametrizations studied, in the one with the smaller strange vector to vector quark coupling ratio, the core-crust thermal coupling (indicated by the sudden drop in the surface temperature) occurs at later times. This indicates agreement with Cas A data, but a more detailed study of the topic will be performed in future investigations.

In the future, we intend to extend our calculations to include relativistic excluded volume effects. Such a consistent approach will provide us with a better idea of the stiffness of the equation of state around deconfinement, and also at higher densities. Work on relativistic versions of excluded volume techniques have been already pursued, for example  in Ref.~\cite{Zhang:1995ux,Zeeb:2002xn,Bugaev:2006pt}, but so far with no guarantee of a physical speed of sound. We also intend to extend our calculation to finite temperature and include neutrino trapping. In this way, we will also be able to study the behavior of strangeness in supernova explosions.

\section*{Aknowledgements}

We thank J. Lattimer, M. Alford and S. Han for fruitful discussions on the physics of compact stars and possible observables. V. D. acknowledges financial support from Helmholtz International Center for FAIR. R.N. acknowledges financial support from CAPES and CNPQ.

\bibstyle{apsrev4-1}
\bibliography{apssamp}% Produces the bibliography via BibTeX.

\end{document}